\documentclass[rsi,twocolumn,nobibnotes]{revtex4-1}
\usepackage{graphicx}
\usepackage{multirow,bigdelim}
\usepackage{bm}
\usepackage{amsmath}
\usepackage{lipsum}
\usepackage[breaklinks=true]{hyperref}

\hypersetup{colorlinks=true,
pdftitle={RPi Beam Profiler},
pdfauthor={James Keaveney},
linkcolor=blue,
citecolor=blue,
urlcolor=blue}

\begin{document}

\title{Automated translating beam profiler for in-situ laser beam spot-size and focal position measurements}
\date{\today}

\author{James Keaveney}
\email{james.keaveney@durham.ac.uk}
\affiliation{Joint Quantum Centre (JQC) Durham-Newcastle, Department of Physics, Durham University, South Road, Durham, DH1 3LE, United Kingdom}

\begin{abstract}
We present a low-cost, high-resolution solution for automated laser-beam profiling with axial translation. The device is based on a Raspberry Pi, Pi Noir CMOS camera, stepper motor and commercial translation stage. We also provide software to run the device. The CMOS sensor is sensitive over a large wavelength range between 300-1100 nm, and can be translated over 25~mm along the beam axis. The sensor head can be reversed without changing its axial position, allowing for a quantitative estimate of beam overlap with counter-propagating laser beams. Although not limited to this application, the intended use for this device is the automated measurement of the focal position and spot-size of a Gaussian laser beam. We present example data of one such measurement to illustrate device performance.
\end{abstract}

\maketitle


\section{Introduction}
\vspace{-0.2cm}

There are a vast number of situations in experimental atomic and optical physics where precise measurement of laser beam profiles is required, from measurements of absolute beam intensity in determining atomic transition dipole matrix elements~\cite{Whiting2016a}, evaluating trap depths in optical dipole traps~\cite{Grimm2000} and laser beam shaping applications~\cite{Dickey2014}, amongst others.
For measuring the focal spot-size of a beam in one dimension, the traditional method is the knife-edge technique~\cite{Suzaki1975,Khosrofian1983,Plass1997,DeAraujo2009}, where the relative transmission is recorded as a sharp edge is moved through the cross-section of the beam. The knife-edge method has a resolution that depends on the quality of the knife-edge and can be as low as 1~$\mu$m~\cite{Dickey2014}, but finding the focal position using this method is a very time consuming and repetitive task, especially if more than one beam axis is to be measured. 
A labour-saving alternative is to use a CMOS image sensor to map the 2D spatial profile of the laser beam. This approach is widely used; there have been recent demonstrations using a webcam~\cite{Langer2013} or the camera built into a smart phone device for this application~\cite{Hossain2015}.

However, for a focussed laser beam one often needs to measure both the beam size at the focus and the axial position of the focus, which necessitates translating the camera or knife-edge along the axial dimension and repeating the profile measurements, which quickly becomes laborious if done manually.
Commercial translating beam profilers are available for this purpose, but are often prohibitively expensive (many thousands of US dollars) and relatively bulky. A recent novel approach used a spatial light modulator to negate the need for any physical translation of the beam~\cite{Schulze2012}, but this was only demonstrated for relatively large (of order 1~mm) beam sizes.

Here we present a device based on a translating CMOS sensor head that uses a Raspberry Pi computer, the Pi Noir CMOS camera, and an inexpensive commercial translation stage. In addition to the hardware, we have developed a computer program to run the image acquisition and analysis.

Apart from the significant cost-saving over current commercial alternatives, our design has two main advantages: first, the small sensor head is the only part of the instrument that is located in the beam path - the main body of the device is relatively small (footprint 245 x 85 mm) and sits adjacent to the beam path, which facilitates in-situ use of the device for most applications. Second, the sensor direction can be reversed without changing the axial position of the sensor (within machining tolerances), allowing for precise determination of the overlap between two beams, which finds use in many optical systems with overlapping counter-propagating beams, for example, in electromagnetically induced transparency~\cite{Gea-Banacloche1995,Noh2012} or four-wave mixing experiments~\cite{Lee2016d,Whiting2016a,Whiting2017}.
In addition, the pixel size of the sensor is smaller than many current commercial systems, though this comes at the cost of a smaller total sensor area; this device is therefore more suitable in the measurement of relatively small beams.
All of the computer software, hardware drawings, CAD files and electronics schematics including a bill-of-materials are available on the GitHub repository for the device~\footnote{\url{https://github.com/jameskeaveney/RPi-Beam-Profiler}}.

\vspace{-0.3cm}
\section{Methods}
\vspace{-0.3cm}
\subsection{Hardware}
\vspace{-0.2cm}

\begin{figure}[t]
\includegraphics[width=1\columnwidth]{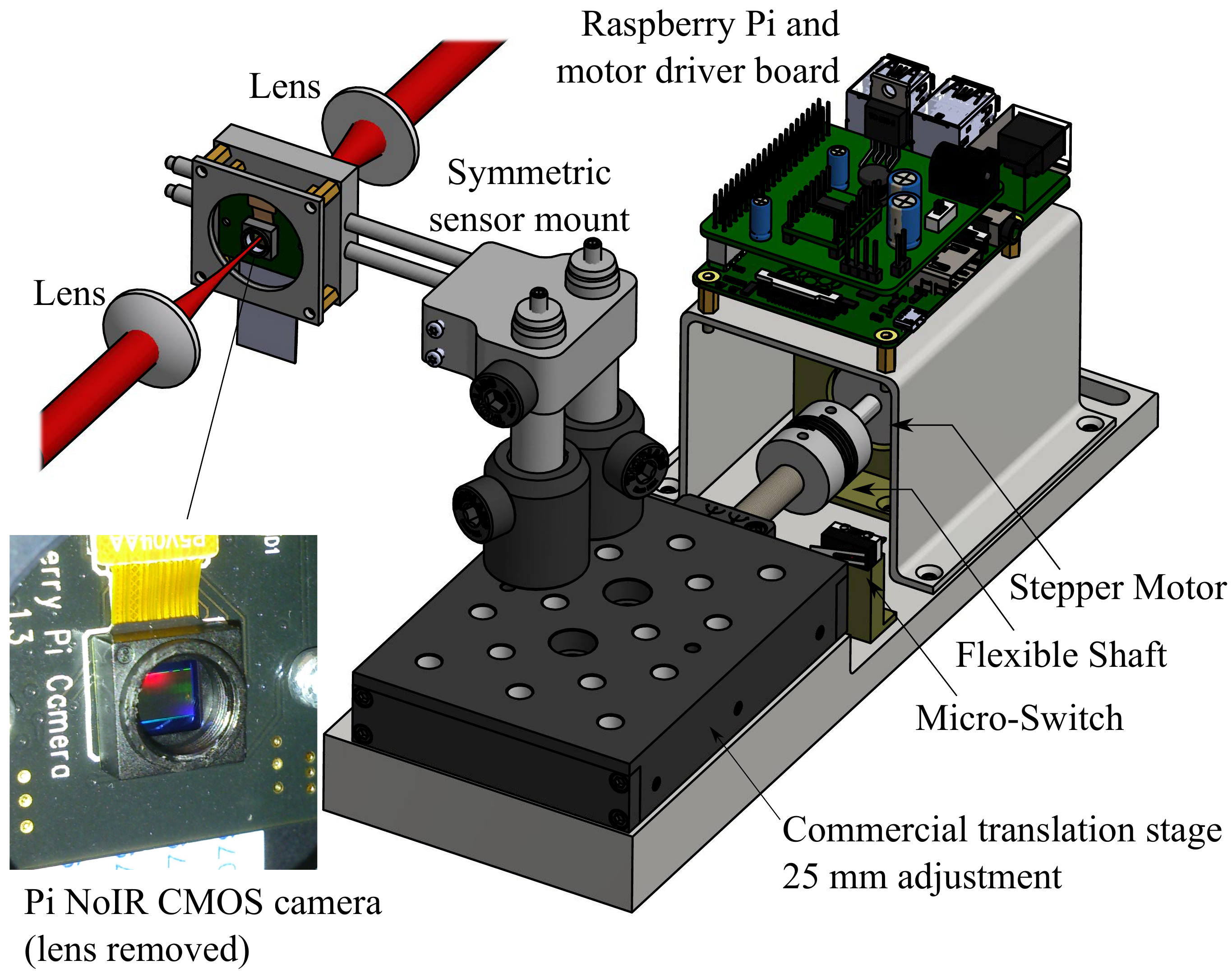}
\caption{Schematic illustration of the beam profiler assembly, showing a typical use scenario where the sensor head is placed between two short focal length lenses (50 mm in this case). See main text for details of the highlighted components. Connecting cables are not shown for clarity. The photo shows the bare image sensor after the original lens has been removed.}
\label{fig:1}
\end{figure}

Figure~\ref{fig:1} shows a rendering of the device.
The camera is mounted on a plate and attached to the translation stage with standard opto-mechanical hardware (part numbers provided in the GitHub repository), allowing for coarse placement in the $xy$ plane. 
To allow for direct profiling of counter-propagating beams, for precise alignment of the relative focal planes, the camera can be placed facing either forwards or backwards along the translation axis. In either orientation the mounts are designed symmetrically around the image-sensor plane, such that the sensor lies at the same axial coordinate, to within machinable precision.
The camera attaches to the Raspberry pi (version 1 model B+, or version 2 or 3 model B)  through a ribbon cable, allowing it to move freely (the standard cable is a little short for this application; several online retailers sell longer cables at little cost). 
A stepper motor controls the movement of the translation stage along the $z$-axis. 
The motor and stage are attached via a custom fine-adjustment screw and flexible shaft coupler, to allow for a small mismatch between the axes of the motor and screw thread. Finally, the whole assembly is mounted on a steel base plate with dimensions $245\times85$~mm.
Control of the stepper motor is via a Pololu driver board based on the DRV8825 controller IC, and is attached to the GPIO pins of the Raspberry Pi.
The angular resolution of the stepper motor and the pitch of the screw thread sets a limit to the axial precision of the device. 
In our case, with a 0.5~mm pitch screw thread and a stepper motor with 0.9~degree resolution (Nanotec ST4209S1006-B), the axial resolution limit is 1.25~$\mu$m. 
The maximum travel of the translation stage is 25~mm, and the step size can be set in software. %
When the profiler software starts, the zero-position is calibrated by moving the translation stage until a microswitch is actuated.
In order to expose the bare sensor, the Pi NoIR CMOS camera has its lens removed (see inset photograph). 
The smallest and largest beam sizes that can be detected are set by the total area of the sensor and the size of a single pixel. 
There are two versions of the PiNoIR camera: version 1 (pre-2016) uses the OmniVision OV5647 sensor with 2592 x 1944 pixels (pixel size 1.4 x 1.4 $\mu$m), whereas version 2 (April 2016 onwards) uses the Sony IMX219 sensor with 3280 x 2464 pixels (pixel size 1.12 x 1.12 $\mu$m). 
Both sensors share the same bit depth of 10-bits and physical sensor size of 3.67 x 2.74 mm. 
We have succesfully tested the device with both sensors, and the software automatically detects which sensor is connected.

To use the device, the only requirements are a suitable power supply (8 - 12~V DC, capable of supplying $> 2$~A), a computer monitor attached to the HDMI port of the Raspberry Pi and a mouse and keyboard attached via USB. All processing is done on the Raspberry Pi. A USB memory stick or similar can be used to extract the gathered data.
%
All custom-made components have relatively simple form, amenable to manufacture in any reasonably-equipped workshop.

\vspace{-0.3cm}
\subsection{Software}
\vspace{-0.3cm}

The beam profiler program can be set up to automatically start when the Raspberry Pi powers on. The analysis program has a graphical user interface and is written in python.
Installation instructions and the source code are provided on the main GitHub page for the beam profiler [link].
\begin{figure}[t]
\includegraphics[width=0.98\columnwidth]{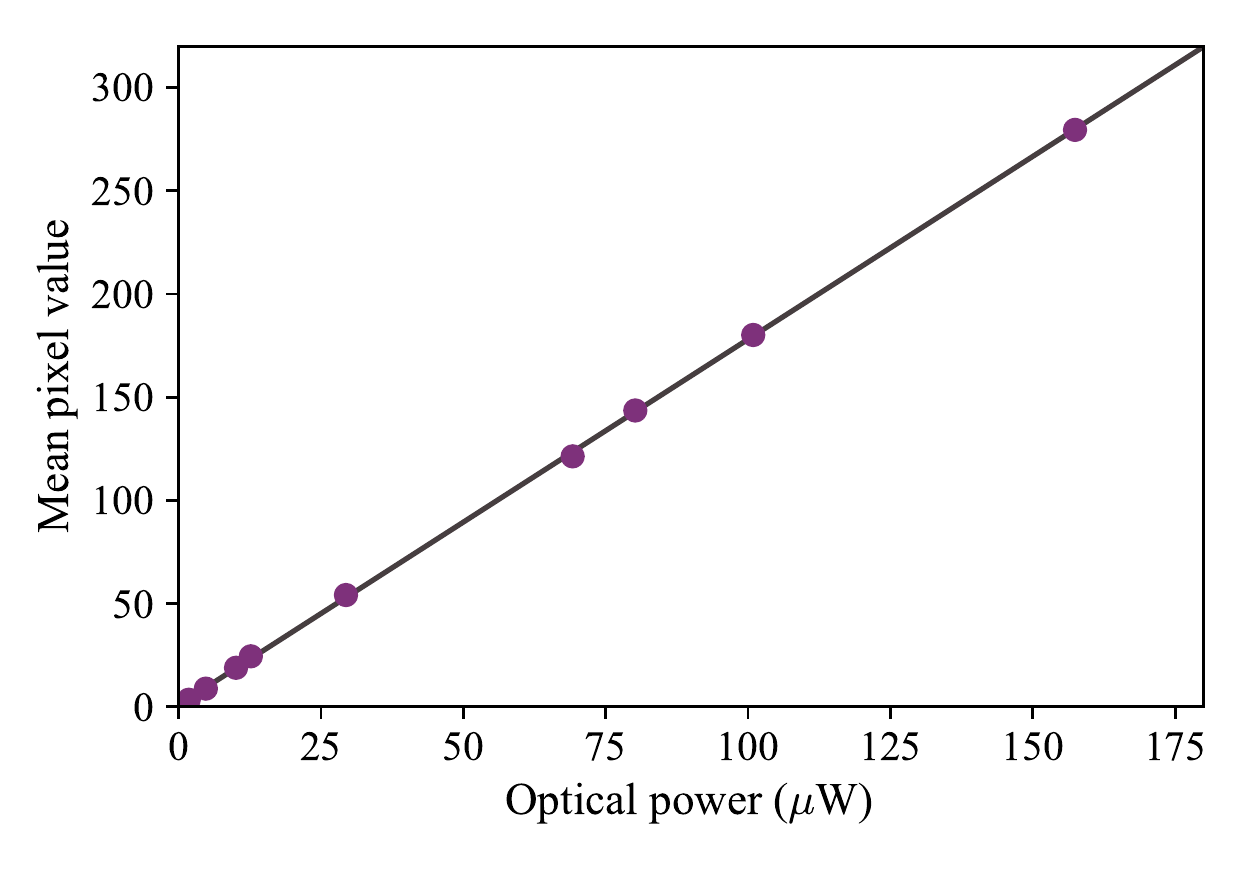}
\caption{Sensor linearity data for the version 2 sensor. Beam width approximately 2 mm, exposure time 85~$\mu$s. 
The fit is to a linear function with the intercept set to zero. The statistical error bars on the points are smaller than the data markers.}
\label{fig:2}
\end{figure}

For alignment purposes, the `Live View' mode can be enabled which uses the camera in video mode. However, to avoid the automatic image enhancement techniques commonly used in digital photography which would artificially alter the data in an unknown way, the raw sensor data must be accessed. 
Since the sensor is a color camera, the raw pixels have color filters arranged in a Bayer pattern (alternating rows of red/green and green/blue pixels). Raw images are captured from the Bayer data of the sensor, and processed into a 2D array. 
From this array, there are software options to interpolate the data back to the full resolution of the sensor (demosaic), or alternatively extract just the red-, green- or blue-filtered pixels, which therefore increases the effective pixel separation by a factor of 2. In principle this resolution decrease can be avoided without interpolation by use of a suitable correction factor~\cite{Langer2013}, but since this has to be empirically determined for each wavelength of light we have not implemented it here.
Image exposure time can be adjusted in software, between a minimum of 9 or 12~$\mu$s (v1 and v2 sensors, respectively) to over 1 second.
A background frame can be measured and subtracted from subsequent images. Once this is removed, the pixel response increases linearly with incident light intensity with near-zero offset, as shown in figure~\ref{fig:2}. The sensor behaves linearly until individual pixels start to saturate, therefore the auto-exposure routine we provide in the software acquires images and adjusts exposure time until the maximum value of any single pixel on an image is just below saturation, to avoid any issues with non-linearity.

The user selects the range and resolution of the translation ($z$-axis) scan, and images are acquired at each position. After each image is acquired, the program integrates over each axis and fits Gaussian functions to the $x$ and $y$ axes separately, extracting the $1/e^2$ radii $w_x$ and $w_y$. This assumes that for an elliptical beam the principal axes of the ellipse are aligned with the camera axes. For more complex beam analysis, the user can choose to automatically save the image data at each $z$ position and manually analyse the data. With the model 3 Raspberry Pi, each data point takes around 30 seconds to move the translation stage, acquire, process and save the image, so a full scan can be completed in approximately 20 minutes depending on the selected translation axis resolution.
After the scan is complete, the axial widths $w_{x,y}(z)$ are fitted to the form of a focussed Gaussian beam given by
\begin{equation}
w(z) = w_0 \sqrt{1 + (z-z_0)^2/z_{\rm R}^2},
\label{eq:WZ}
\end{equation}
where $w_0$ is the beam waist, $z_0$ the focal position, and $z_{\rm R}=\pi w_0^2 / \lambda$ is the Rayleigh range with $\lambda$ the wavelength of light. However, this assumes a beam-quality factor $M^2 = 1$ (i.e. a perfect TEM$_{00}$ mode) which is rarely the case experimentally, so in practice we let $z_{\rm R}$ be a floating parameter in addition to $w_0$ and $z_0$. If the beam profile is Gaussian, then $M^2$ can be extracted by assuming that $z_{\rm R}=\pi w_0^2 / M^2 \lambda$~\cite{Johnston1998}.
It is possible to implement a full $M^2$ measurement for non-Gaussian beams using this system with suitable changes to the image analysis routine; this woud require using the ISO standard~\cite{ISO11146-1} second-moment $D4\sigma$ beam width measurement instead of simple Gaussian fitting. However, since this is beyond our typical use case, we have not implemented it here.


\begin{figure}[t]
\includegraphics[width=0.98\columnwidth]{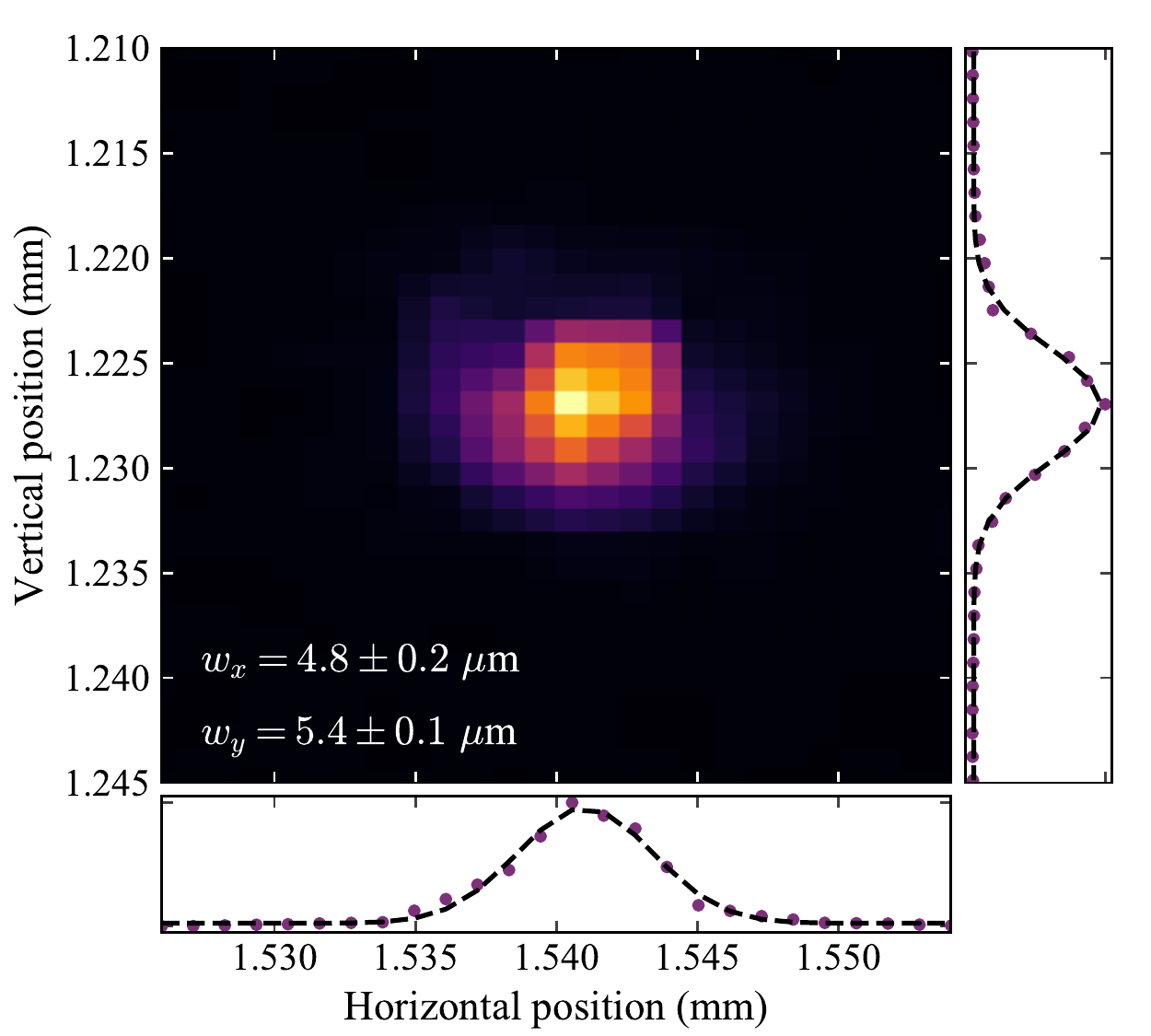}
\caption{Exampe image showing a tightly focussed beam with a relatively small spot size. Integrating over each dimension of the image separately we extract and fit Gaussian profiles, shown on the bottom and right panels, with $1/e^2$ radii as shown on the figure. The purple points are the data, dashed black lines are the best fit lines.}
\label{fig:3}
\end{figure}

\vspace{-0.3cm}
\section{Results}
\vspace{-0.2cm}

In the remainder of the paper, we illustrate the use of the beam profiler through example data.

\begin{figure}[t]
\includegraphics[width=0.98\columnwidth]{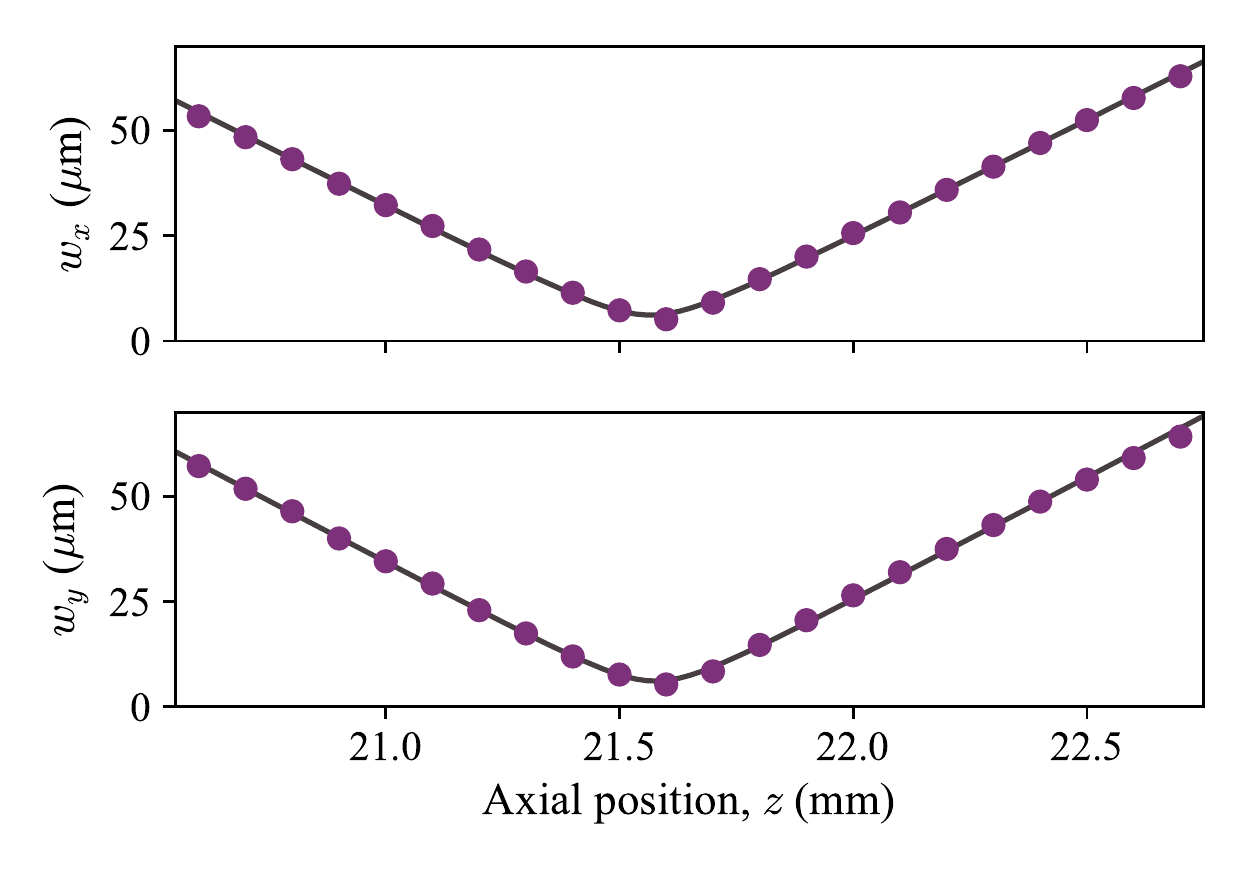}
\caption{Beam waists $w_x$ and $w_y$ as a function of axial position. Purple points are data; the statistical error bars from the individual fits to each image are smaller than the data point markers. The solid black lines are fits to eq.~(1).}
\label{fig:4}
\end{figure}

A diode laser with 780~nm wavelength was coupled through a single-mode optical fiber. After the fiber, the beam was collimated to a 1/e$^2$ radius of approximately 0.8~mm, and then incident on a single-element aspheric lens with 15~mm focal length. The damage threshold was not explicitly tested, but to avoid damaging the sensor, we kept the incident optical power low ($< 1$~mW) using neutral density filters. The sensor saturates (even at the minimum exposure time) at a much lower intensity than damage occurs, therefore in practice the power can be set by coarsely placing the beam profiler near the focus of the beam and adjusting the incident intensity such that the sensor is not saturating on any single pixel with the lowest exposure time.
Figure~\ref{fig:3} shows an example image taken with the v2 sensor.
The image is taken at the focal position of a 15~mm focal length aspheric lens, so the focal spot is only a few microns. 
We show the (cropped) image data and the integrated horizontal and vertical profiles, along with the fits to Gaussians. We extract radii $w_x = 4.8 \pm 0.2$~$\mu$m and $w_y = 5.4\pm0.1$~$\mu$m. These waists are consistent with the 5.0~$\mu$m that we expect from simple Gaussian beam propagation~\cite{Brooker2003} assuming a perfectly collimated input beam. The fit uses a standard Levenburg-Marquardt routine and the errors in the fit parameters are estimated using the square-root of the diagonal elements of the covariance matrix~\cite{Hughes2010}.
In figure~\ref{fig:4} we show an example of $x$- and $y$-axis beam waists as a function of translation distance. In this case, the selected axial resolution is 100~$\mu$m. After running a translation scan, the software fits the processed beam width data to equation~\ref{eq:WZ}. The result of the fits are shown by the black lines; the $x$ and $y$ axes are fitted independently and both are clearly in agreement with the data. In this case, we find for the x-axis a focal size $w_0 = 6.13 \pm 0.03 \ \mu$m, focal position $z_0 = 21.566 \pm 0.001$~mm, and Rayleigh range $z_{\rm R} = 0.110 \pm 0.001$~mm. For the y-axis, we find $w_0 = 6.11 \pm 0.03 \ \mu$m, focal position $z_0 = 21.577 \pm 0.001$~mm, and Rayleigh range $z_{\rm R} = 0.104 \pm 0.001$~mm. All error bars are taken from the square root of the diagonal elements of the covariance matrix. Assuming that $z_{\rm R} = \pi w_0^2 / M^2 \lambda$, we can estimate the beam quality factor for the two axes as $M^2_x = 1.38 \pm 0.01$ and $M^2_y = 1.44 \pm 0.01$, which is reasonable for this type of laser.

\section{Conclusions}
\vspace{-0.2cm}

In conclusion, we have presented an automatic translating beam profiler that can be used for in-situ measurements of focal spot size and positions. The system has some advantages over current commercial systems, besides the cost reduction; the pixel size is smaller than many CCD-based commercial beam profilers, and the measurement head can be placed into an already existing optical setup, facing in both the $\pm z$ directions allowing for a quantitative estimate of beam overlap when using counter-propagating beams. The software is open-source and the analysis methods could be extended to include the ISO standard second-moment $D4\sigma$ beam width and $M^2$ measurements in future versions.

I gratefully acknowledge Daniel J. Whiting for testing of the device, Ifan G. Hughes for fruitful discussions and EPSRC (grant number EP/L023024/1) for funding. In addition to the GitHub repository~[15], the data presented in this paper are available online at $<$DOI added at proof stage$>$.

\vspace{-0.3cm}
\bibliography{library}

\begin{thebibliography}{19}%
\makeatletter
\providecommand \@ifxundefined [1]{%
 \@ifx{#1\undefined}
}%
\providecommand \@ifnum [1]{%
 \ifnum #1\expandafter \@firstoftwo
 \else \expandafter \@secondoftwo
 \fi
}%
\providecommand \@ifx [1]{%
 \ifx #1\expandafter \@firstoftwo
 \else \expandafter \@secondoftwo
 \fi
}%
\providecommand \natexlab [1]{#1}%
\providecommand \enquote  [1]{``#1''}%
\providecommand \bibnamefont  [1]{#1}%
\providecommand \bibfnamefont [1]{#1}%
\providecommand \citenamefont [1]{#1}%
\providecommand \href@noop [0]{\@secondoftwo}%
\providecommand \href [0]{\begingroup \@sanitize@url \@href}%
\providecommand \@href[1]{\@@startlink{#1}\@@href}%
\providecommand \@@href[1]{\endgroup#1\@@endlink}%
\providecommand \@sanitize@url [0]{\catcode `\\12\catcode `\$12\catcode
  `\&12\catcode `\#12\catcode `\^12\catcode `\_12\catcode `\%12\relax}%
\providecommand \@@startlink[1]{}%
\providecommand \@@endlink[0]{}%
\providecommand \url  [0]{\begingroup\@sanitize@url \@url }%
\providecommand \@url [1]{\endgroup\@href {#1}{\urlprefix }}%
\providecommand \urlprefix  [0]{URL }%
\providecommand \Eprint [0]{\href }%
\providecommand \doibase [0]{http://dx.doi.org/}%
\providecommand \selectlanguage [0]{\@gobble}%
\providecommand \bibinfo  [0]{\@secondoftwo}%
\providecommand \bibfield  [0]{\@secondoftwo}%
\providecommand \translation [1]{[#1]}%
\providecommand \BibitemOpen [0]{}%
\providecommand \bibitemStop [0]{}%
\providecommand \bibitemNoStop [0]{.\EOS\space}%
\providecommand \EOS [0]{\spacefactor3000\relax}%
\providecommand \BibitemShut  [1]{\csname bibitem#1\endcsname}%
\let\auto@bib@innerbib\@empty
\bibitem [{\citenamefont {Whiting}\ \emph {et~al.}(2016)\citenamefont
  {Whiting}, \citenamefont {Keaveney}, \citenamefont {Adams},\ and\
  \citenamefont {Hughes}}]{Whiting2016a}%
  \BibitemOpen
  \bibfield  {author} {\bibinfo {author} {\bibfnamefont {D.~J.}\ \bibnamefont
  {Whiting}}, \bibinfo {author} {\bibfnamefont {J.}~\bibnamefont {Keaveney}},
  \bibinfo {author} {\bibfnamefont {C.~S.}\ \bibnamefont {Adams}}, \ and\
  \bibinfo {author} {\bibfnamefont {I.~G.}\ \bibnamefont {Hughes}},\ }\href
  {\doibase 10.1103/PhysRevA.93.043854} {\bibfield  {journal} {\bibinfo
  {journal} {Phys. Rev. A}\ }\textbf {\bibinfo {volume} {93}},\ \bibinfo
  {pages} {043854} (\bibinfo {year} {2016})},\ \Eprint
  {http://arxiv.org/abs/1602.08944} {arXiv:1602.08944} \BibitemShut {NoStop}%
\bibitem [{\citenamefont {Grimm}\ \emph {et~al.}(2000)\citenamefont {Grimm},
  \citenamefont {Weidem{\"{u}}ller},\ and\ \citenamefont
  {Ovchinnikov}}]{Grimm2000}%
  \BibitemOpen
  \bibfield  {author} {\bibinfo {author} {\bibfnamefont {R.}~\bibnamefont
  {Grimm}}, \bibinfo {author} {\bibfnamefont {M.}~\bibnamefont
  {Weidem{\"{u}}ller}}, \ and\ \bibinfo {author} {\bibfnamefont {Y.~B.}\
  \bibnamefont {Ovchinnikov}},\ }\href {\doibase 10.1016/S1049-250X(08)60186-X}
  {\bibfield  {journal} {\bibinfo  {journal} {Adv. At. Mol. Opt. Phys.}\
  }\textbf {\bibinfo {volume} {42}},\ \bibinfo {pages} {95} (\bibinfo {year}
  {2000})},\ \Eprint {http://arxiv.org/abs/9902072} {arXiv:9902072 [physics]}
  \BibitemShut {NoStop}%
\bibitem [{\citenamefont {Dickey}(2014)}]{Dickey2014}%
  \BibitemOpen
  \bibinfo {editor} {\bibfnamefont {F.~M.}\ \bibnamefont {Dickey}},\ ed.,\
  \href@noop {} {\emph {\bibinfo {title} {{Laser Beam Shaping: Theory and
  Techniques}}}},\ \bibinfo {edition} {2nd}\ ed.\ (\bibinfo  {publisher} {CRC
  Press},\ \bibinfo {year} {2014})\ pp.\ \bibinfo {pages} {1--81}\BibitemShut
  {NoStop}%
\bibitem [{\citenamefont {Suzaki}\ and\ \citenamefont
  {Tachibana}(1975)}]{Suzaki1975}%
  \BibitemOpen
  \bibfield  {author} {\bibinfo {author} {\bibfnamefont {Y.}~\bibnamefont
  {Suzaki}}\ and\ \bibinfo {author} {\bibfnamefont {A.}~\bibnamefont
  {Tachibana}},\ }\href {\doibase 10.1364/AO.14.002809} {\bibfield  {journal}
  {\bibinfo  {journal} {Appl. Opt.}\ }\textbf {\bibinfo {volume} {14}},\
  \bibinfo {pages} {2809} (\bibinfo {year} {1975})}\BibitemShut {NoStop}%
\bibitem [{\citenamefont {Khosrofian}\ and\ \citenamefont
  {Garetz}(1983)}]{Khosrofian1983}%
  \BibitemOpen
  \bibfield  {author} {\bibinfo {author} {\bibfnamefont {J.~M.}\ \bibnamefont
  {Khosrofian}}\ and\ \bibinfo {author} {\bibfnamefont {B.~A.}\ \bibnamefont
  {Garetz}},\ }\href {\doibase 10.1364/AO.22.003406} {\bibfield  {journal}
  {\bibinfo  {journal} {Appl. Opt.}\ }\textbf {\bibinfo {volume} {22}},\
  \bibinfo {pages} {3406} (\bibinfo {year} {1983})}\BibitemShut {NoStop}%
\bibitem [{\citenamefont {Plass}\ \emph {et~al.}(1997)\citenamefont {Plass},
  \citenamefont {Maestle}, \citenamefont {Wittig}, \citenamefont {Voss},\ and\
  \citenamefont {Giesen}}]{Plass1997}%
  \BibitemOpen
  \bibfield  {author} {\bibinfo {author} {\bibfnamefont {W.}~\bibnamefont
  {Plass}}, \bibinfo {author} {\bibfnamefont {R.}~\bibnamefont {Maestle}},
  \bibinfo {author} {\bibfnamefont {K.}~\bibnamefont {Wittig}}, \bibinfo
  {author} {\bibfnamefont {A.}~\bibnamefont {Voss}}, \ and\ \bibinfo {author}
  {\bibfnamefont {A.}~\bibnamefont {Giesen}},\ }\href {\doibase
  10.1016/S0030-4018(96)00527-5} {\bibfield  {journal} {\bibinfo  {journal}
  {Opt. Commun.}\ }\textbf {\bibinfo {volume} {134}},\ \bibinfo {pages} {21}
  (\bibinfo {year} {1997})}\BibitemShut {NoStop}%
\bibitem [{\citenamefont {de~Ara{\'{u}}jo}\ \emph {et~al.}(2009)\citenamefont
  {de~Ara{\'{u}}jo}, \citenamefont {Silva}, \citenamefont {de~Lima},
  \citenamefont {Pereira},\ and\ \citenamefont {de~Oliveira}}]{DeAraujo2009}%
  \BibitemOpen
  \bibfield  {author} {\bibinfo {author} {\bibfnamefont {M.~A.}\ \bibnamefont
  {de~Ara{\'{u}}jo}}, \bibinfo {author} {\bibfnamefont {R.}~\bibnamefont
  {Silva}}, \bibinfo {author} {\bibfnamefont {E.}~\bibnamefont {de~Lima}},
  \bibinfo {author} {\bibfnamefont {D.~P.}\ \bibnamefont {Pereira}}, \ and\
  \bibinfo {author} {\bibfnamefont {P.~C.}\ \bibnamefont {de~Oliveira}},\
  }\href {\doibase 10.1364/AO.48.000393} {\bibfield  {journal} {\bibinfo
  {journal} {Appl. Opt.}\ }\textbf {\bibinfo {volume} {48}},\ \bibinfo {pages}
  {393} (\bibinfo {year} {2009})}\BibitemShut {NoStop}%
\bibitem [{\citenamefont {Langer}\ \emph {et~al.}(2013)\citenamefont {Langer},
  \citenamefont {Hochreiner}, \citenamefont {Burgholzer},\ and\ \citenamefont
  {Berer}}]{Langer2013}%
  \BibitemOpen
  \bibfield  {author} {\bibinfo {author} {\bibfnamefont {G.}~\bibnamefont
  {Langer}}, \bibinfo {author} {\bibfnamefont {A.}~\bibnamefont {Hochreiner}},
  \bibinfo {author} {\bibfnamefont {P.}~\bibnamefont {Burgholzer}}, \ and\
  \bibinfo {author} {\bibfnamefont {T.}~\bibnamefont {Berer}},\ }\href
  {\doibase 10.1016/j.optlaseng.2012.12.010} {\bibfield  {journal} {\bibinfo
  {journal} {Opt. Lasers Eng.}\ }\textbf {\bibinfo {volume} {51}},\ \bibinfo
  {pages} {571} (\bibinfo {year} {2013})}\BibitemShut {NoStop}%
\bibitem [{\citenamefont {Hossain}\ \emph {et~al.}(2015)\citenamefont
  {Hossain}, \citenamefont {Canning}, \citenamefont {Cook},\ and\ \citenamefont
  {Jamalipour}}]{Hossain2015}%
  \BibitemOpen
  \bibfield  {author} {\bibinfo {author} {\bibfnamefont {M.~A.}\ \bibnamefont
  {Hossain}}, \bibinfo {author} {\bibfnamefont {J.}~\bibnamefont {Canning}},
  \bibinfo {author} {\bibfnamefont {K.}~\bibnamefont {Cook}}, \ and\ \bibinfo
  {author} {\bibfnamefont {A.}~\bibnamefont {Jamalipour}},\ }\href {\doibase
  10.1364/OL.40.005156} {\bibfield  {journal} {\bibinfo  {journal} {Opt.
  Lett.}\ }\textbf {\bibinfo {volume} {40}},\ \bibinfo {pages} {5156} (\bibinfo
  {year} {2015})}\BibitemShut {NoStop}%
\bibitem [{\citenamefont {Schulze}\ \emph {et~al.}(2012)\citenamefont
  {Schulze}, \citenamefont {Flamm}, \citenamefont {Duparr{\'{e}}},\ and\
  \citenamefont {Forbes}}]{Schulze2012}%
  \BibitemOpen
  \bibfield  {author} {\bibinfo {author} {\bibfnamefont {C.}~\bibnamefont
  {Schulze}}, \bibinfo {author} {\bibfnamefont {D.}~\bibnamefont {Flamm}},
  \bibinfo {author} {\bibfnamefont {M.}~\bibnamefont {Duparr{\'{e}}}}, \ and\
  \bibinfo {author} {\bibfnamefont {A.}~\bibnamefont {Forbes}},\ }\href
  {\doibase 10.1364/OL.37.004687} {\bibfield  {journal} {\bibinfo  {journal}
  {Opt. Lett.}\ }\textbf {\bibinfo {volume} {37}},\ \bibinfo {pages} {4687}
  (\bibinfo {year} {2012})}\BibitemShut {NoStop}%
\bibitem [{\citenamefont {Gea-Banacloche}\ \emph {et~al.}(1995)\citenamefont
  {Gea-Banacloche}, \citenamefont {Li}, \citenamefont {Jin},\ and\
  \citenamefont {Xiao}}]{Gea-Banacloche1995}%
  \BibitemOpen
  \bibfield  {author} {\bibinfo {author} {\bibfnamefont {J.}~\bibnamefont
  {Gea-Banacloche}}, \bibinfo {author} {\bibfnamefont {Y.-Q.}\ \bibnamefont
  {Li}}, \bibinfo {author} {\bibfnamefont {S.-Z.}\ \bibnamefont {Jin}}, \ and\
  \bibinfo {author} {\bibfnamefont {M.}~\bibnamefont {Xiao}},\ }\href {\doibase
  10.1103/PhysRevA.51.576} {\bibfield  {journal} {\bibinfo  {journal} {Phys.
  Rev. A}\ }\textbf {\bibinfo {volume} {51}},\ \bibinfo {pages} {576} (\bibinfo
  {year} {1995})}\BibitemShut {NoStop}%
\bibitem [{\citenamefont {Noh}\ and\ \citenamefont {Moon}(2012)}]{Noh2012}%
  \BibitemOpen
  \bibfield  {author} {\bibinfo {author} {\bibfnamefont {H.-R.}\ \bibnamefont
  {Noh}}\ and\ \bibinfo {author} {\bibfnamefont {H.~S.}\ \bibnamefont {Moon}},\
  }\href {\doibase 10.1103/PhysRevA.85.033817} {\bibfield  {journal} {\bibinfo
  {journal} {Phys. Rev. A}\ }\textbf {\bibinfo {volume} {85}},\ \bibinfo
  {pages} {033817} (\bibinfo {year} {2012})}\BibitemShut {NoStop}%
\bibitem [{\citenamefont {Lee}\ \emph {et~al.}(2016)\citenamefont {Lee},
  \citenamefont {Lee}, \citenamefont {Kim},\ and\ \citenamefont
  {Moon}}]{Lee2016d}%
  \BibitemOpen
  \bibfield  {author} {\bibinfo {author} {\bibfnamefont {Y.-S.}\ \bibnamefont
  {Lee}}, \bibinfo {author} {\bibfnamefont {S.~M.}\ \bibnamefont {Lee}},
  \bibinfo {author} {\bibfnamefont {H.}~\bibnamefont {Kim}}, \ and\ \bibinfo
  {author} {\bibfnamefont {H.~S.}\ \bibnamefont {Moon}},\ }\href {\doibase
  10.1364/OE.24.028083} {\bibfield  {journal} {\bibinfo  {journal} {Opt.
  Express}\ }\textbf {\bibinfo {volume} {24}},\ \bibinfo {pages} {28083}
  (\bibinfo {year} {2016})},\ \Eprint {http://arxiv.org/abs/1609.02378}
  {arXiv:1609.02378} \BibitemShut {NoStop}%
\bibitem [{\citenamefont {Whiting}\ \emph {et~al.}(2017)\citenamefont
  {Whiting}, \citenamefont {{\v{S}}ibali{\'{c}}}, \citenamefont {Keaveney},
  \citenamefont {Adams},\ and\ \citenamefont {Hughes}}]{Whiting2017}%
  \BibitemOpen
  \bibfield  {author} {\bibinfo {author} {\bibfnamefont {D.~J.}\ \bibnamefont
  {Whiting}}, \bibinfo {author} {\bibfnamefont {N.}~\bibnamefont
  {{\v{S}}ibali{\'{c}}}}, \bibinfo {author} {\bibfnamefont {J.}~\bibnamefont
  {Keaveney}}, \bibinfo {author} {\bibfnamefont {C.~S.}\ \bibnamefont {Adams}},
  \ and\ \bibinfo {author} {\bibfnamefont {I.~G.}\ \bibnamefont {Hughes}},\
  }\href {\doibase 10.1103/PhysRevLett.118.253601} {\bibfield  {journal}
  {\bibinfo  {journal} {Phys. Rev. Lett.}\ }\textbf {\bibinfo {volume} {118}},\
  \bibinfo {pages} {253601} (\bibinfo {year} {2017})},\ \Eprint
  {http://arxiv.org/abs/1612.05467} {arXiv:1612.05467} \BibitemShut {NoStop}%
\bibitem [{Note1()}]{Note1}%
  \BibitemOpen
  \bibinfo {note} {\protect \url
  {https://github.com/jameskeaveney/RPi-Beam-Profiler}}\BibitemShut {NoStop}%
\bibitem [{\citenamefont {Johnston}(1998)}]{Johnston1998}%
  \BibitemOpen
  \bibfield  {author} {\bibinfo {author} {\bibfnamefont {T.~F.}\ \bibnamefont
  {Johnston}},\ }\href {\doibase 10.1364/AO.37.004840} {\bibfield  {journal}
  {\bibinfo  {journal} {Appl. Opt.}\ }\textbf {\bibinfo {volume} {37}},\
  \bibinfo {pages} {4840} (\bibinfo {year} {1998})}\BibitemShut {NoStop}%
\bibitem [{ISO(2005)}]{ISO11146-1}%
  \BibitemOpen
  \href {https://www.iso.org/standard/33625.html} {\enquote {\bibinfo {title}
  {{ISO 11146-1:2005(en) and 11146-2:2005(en) Lasers and laser-related
  equipment — Test methods for laser beam widths, divergence angles and beam
  propagation ratios — Parts 1 and 2}},}\ } (\bibinfo {year}
  {2005})\BibitemShut {NoStop}%
\bibitem [{\citenamefont {Brooker}(2003)}]{Brooker2003}%
  \BibitemOpen
  \bibfield  {author} {\bibinfo {author} {\bibfnamefont {G.}~\bibnamefont
  {Brooker}},\ }\href@noop {} {\emph {\bibinfo {title} {{Modern Classical
  Optics}}}}\ (\bibinfo  {publisher} {OUP},\ \bibinfo {address} {Oxford},\
  \bibinfo {year} {2003})\BibitemShut {NoStop}%
\bibitem [{\citenamefont {Hughes}\ and\ \citenamefont
  {Hase}(2010)}]{Hughes2010}%
  \BibitemOpen
  \bibfield  {author} {\bibinfo {author} {\bibfnamefont {I.~G.}\ \bibnamefont
  {Hughes}}\ and\ \bibinfo {author} {\bibfnamefont {T.~P.~A.}\ \bibnamefont
  {Hase}},\ }\href@noop {} {\emph {\bibinfo {title} {{Measurements and their
  Uncertainties: A Practical Guide to Modern Error Analysis}}}}\ (\bibinfo
  {publisher} {OUP},\ \bibinfo {address} {Oxford},\ \bibinfo {year}
  {2010})\BibitemShut {NoStop}%
\end{thebibliography}%

\end{document}